# An upgraded ultra-high vacuum magnetron-sputtering system for high-versatility and software-controlled deposition

Arnaud le Febvrier[1*], Ludvig Landälv[1,2], Thomas Liersch[1], David Sandmark[1,3], Per Sandström[1], Per Eklund[1*]

**Affiliations:**

[1]*Thin Film Physics Division, Department of Physics, Chemistry and Biology (IFM), Linköping University, SE-58183 Linköping, Sweden*

[2]*Present address: Sandvik Coromant AB, Stockholm SE-12680, Sweden*

[3]*Present address: Saab AB Aeronautics, Bröderna Ugglas gata, 582 54 Linköping*

**\*Corresponding author:** arnaud.le.febvrier@liu.se ; per.eklund@liu.se

**Abstract:** Magnetron sputtering is a widely used physical vapor deposition technique. Reactive sputtering is used for the deposition of, e.g, oxides, nitrides and carbides. In fundamental research, versatility is essential when designing or upgrading a deposition chamber. Furthermore, automated deposition systems are the norm in industrial production, but relatively uncommon in laboratory-scale systems used primarily for fundamental research. Combining automatization and computerized control with the required versatility for fundamental research constitutes a challenge in designing, developing, and upgrading laboratory deposition systems. The present article provides a detailed description of the design of a lab-scale deposition chamber for magnetron sputtering used for the deposition of metallic, oxide, nitride and oxynitride films with automated controls, dc or pulsed bias, and combined with a coil to enhance the plasma density near the substrate. LabVIEW software (provided as Supplementary Information) has been developed for a high degree of computerized or automated control of hardware and processes control and logging of process details.

# 1. Introduction

Magnetron sputtering is a versatile and widely used physical vapor deposition (PVD) technique [1]. It can be operated in a wide range of discharge modes [2] such as direct current (dc), pulsed-dc, radio-frequency (rf), and ionized sputter-deposition techniques [3, 4] like high-power impulse magnetron sputtering (HiPIMS)[5]. At the substrate position, a bias voltage can be applied to control the bombardment and energy of the incident species and modify the film growth [1, 6, 7]. Reactive sputtering, where reactive gases are introduced, can be used for the deposition of, e.g, oxides, nitrides and carbides [8-10]. For these purposes, especially in fundamental research, versatility is a key point when designing or upgrading a deposition chamber. Furthermore, largely or fully automated deposition systems are the norm in industrial production, but relatively uncommon in laboratory-scale systems used primarily for fundamental research. Combining largely automated and computerized control with the required versatility for fundamental research constitutes a challenge in designing, developing, and upgrading laboratory deposition systems.

The present article provides a detailed description of the design of a lab-scale deposition chamber for magnetron sputtering used for the deposition of metallic, oxide, nitride and oxynitride films with automated controls, dc or pulsed bias, and combined with a coil to enhance the plasma density near the substrate[11, 12]. Software has been developed (LabVIEW) for a high degree of computerized or automated control of hardware and processes control and logging of process details.

The deposition chamber described here was originally acquired in 2004, and has been used for a wide range of materials, such as MAX-phases [13], mixed transition metal oxides for hard-coating and cutting-tool applications (such as (Zr,Al)$_2$O$_3$ [14, 15], (Cr,Al)$_2$O$_3$ [16, 17] and AlVO [18]), piezoelectric (Sc,Al)N [19], thermoelectric nitrides and oxides [20-22], and transparent hard oxynitride coatings [23]. As of 2019, large upgrades and additions were implemented. The purpose of the present article is to provide a detailed description of the system and the design considerations.

## 2. Deposition chamber

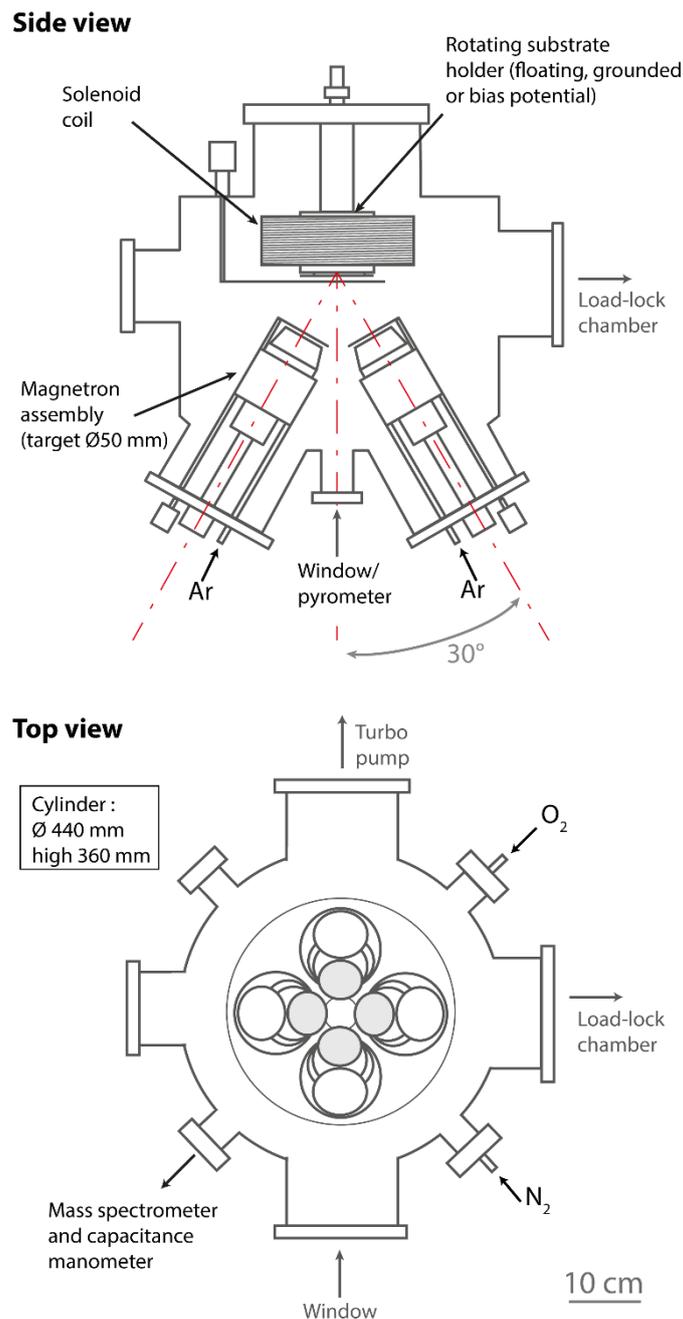

**Figure 1:** Schematic top view and side view of the main chamber of the deposition chamber.

In summary, the deposition system is equipped with four magnetron sputtering sources to which dc, rf, or pulsed-dc voltage can be applied. The ultrahigh vacuum (UHV) chamber has with a base pressure typically lower than $1.3\times10^{-7}$ Pa ($1\times10^{-9}$ Torr). Gas inlets are available for argon, oxygen, and nitrogen. Those three gases, in combination with the different power supplies to the magnetrons, enable the deposition of metal, oxide, nitride and oxynitride films. The sample stage can be heated up to 1100°C (substrate temperature). A bias voltage (dc, pulsed dc, or rf) can be applied to the substrate during deposition and can be combined with a coil to enhance the plasma density near the substrate. The deposition system is largely automated and computer-controlled, allowing improved reproducibility with process control and logging. The high level of automation also enables complex deposition sequences.

Figure 1 is a schematic view of the main chamber. The deposition system comprises a main chamber made of low carbon stainless steel connected to a load-lock chamber for loading with a transfer arm. The load-lock chamber is connected to the pumping system (turbomolecular pump (Varian V70 with nominal pumping speed of 68 l/s ($N_2$)) + dry screw backing pump (Ebara EV-PA 250 with nominal pumping speed of 230 l/s ($N_2$)) consisting of two separated paths for pumping at high pressure range (atmospheric pressure to $10^{-1}$ Pa ($10^{-3}$ Torr)) and at low pressure range (from $10^{-1}$ Pa ($10^{-3}$ Torr) to $10^{-6}$ Pa ($10^{-8}$ Torr)). The minimum pressure in the load-lock chamber is about $1.3 \times 10^{-6}$ Pa ($1 \times 10^{-8}$ Torr). The main chamber (Figure. 2) with an approximate volume of 0.55 m³ (cylinder: diameter of 0.44 m and high of 0.36 m) is connected to a turbomolecular pump with a dry screw backing pump (Ebara EV-PA 250 with nominal pumping speed of 230 l/s ($N_2$)) maintaining a foreline pressure lower than 0.03 Pa ($2 \times 10^{-4}$ Torr) during normal operation. Typically, a base pressure lower than $< 3.9 \times 10^{-7}$ Pa ($< 3 \times 10^{-9}$ Torr) is reached after baking. The turbomolecular pump (Agilent V551) with a nominal pumping speed of 550 l/s ($N_2$) was replaced in 2020 by another turbomolecular pump (Agilent TwisTorr 704 FS) with a nominal pumping speed of 650 l/s ($N_2$). The baking is performed using electric band heater covering the chamber wall. Typical baking parameters are a temperature of 140°C for 48h. The chamber is also equipped with a quadrupole mass spectrometer (MKS HPQ-2) for analyzing residual gases during or after baking. In order to minimize the outgassing during deposition of materials at elevated temperature, the chamber walls are water-cooled and kept near room temperature.

The pressures are monitored by a Granville-Philips ion gauge at low pressure and by an MKS Baratron® capacitance manometer (connected to MKS PDR2000) during deposition (pressure range of 0.01-50 mTorr with a resolution of 0.01 mtorr (range of $1 \times 10^{-3}$ - 6 Pa with a resolution of 0.001 Pa)). In common use, the maximum pumping speed is used during deposition. In some specific cases, a throttle valve can be used to restrict the effective pumping speed of the turbomolecular pump allowing more freedom on adjusting pressure versus flow of gases.

## 3. Gas inlets / deposition pressures

The deposition chamber is connected to argon (purity 99.999 %), nitrogen (purity 99.999 %), and oxygen (purity 99.999 %) inlet. The argon and nitrogen gas lines are equipped with purifiers. The different gas flows are controlled through an MKS Multigas Controller 647B. Different mass flow controllers (MFC) are used to control the gas flows: a 100 sccm MFC for Ar, a 100 sccm MFC for $N_2$ and two MFC of 10 sccm and 50 sccm for $O_2$. The nitrogen and the oxygen inlets are situated on the wall of the chamber, while the argon gas inlet is distributed equally after the MFC on each magnetron assembly with individual valves. The partial pressures of the different gases used for deposition can be monitored with the quadrupole mass spectrometer connected to the chamber. At typical deposition pressures of 0.1 - 0.8 Pa (1 - 6 mTorr), the effective pumping speed is reduced to approximately half the nominal pumping speed value [8]. For example, using the Agilent V551 (replaced in 2020, see section 2), an effective pumping speed of 216 l/s was estimated for 0.33 Pa (2.5 mTorr) of argon pressure (45 sccm of Ar flow without throttle valve). These relatively high effective pumping speeds with the volume of the chamber is an advantage when depositing metal under reactive gas where no or small hysteresis has been observed. Strijckmans *et al.* detailed the different solutions in order to reduce a hysteresis observed on metal target with reactive gases [8]. The chamber is also equipped with a quadrupole mass spectrometer (MKS HPQ-2) for analyzing residual gases during or after baking and the possibility to monitor of the partial pressures of reactive gases in the chamber during deposition.

## 4. Magnetron assembly

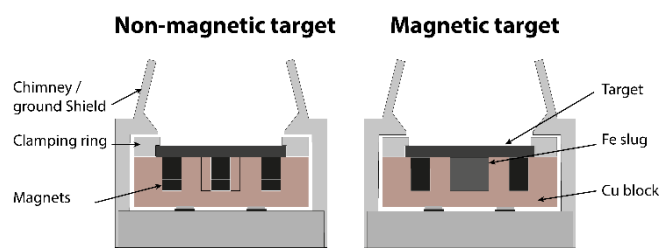

**Figure 2:** Schematic view of the magnetron assembly with two different magnet configurations intended for use with a regular non-magnetic target (left) and for use with a magnetic target (right). In the latter case, note that the central magnets are removed and replaced by an iron slug.

The chamber is equipped with four magnetron assemblies (AJA-A320) in a close-field unbalanced configuration with grounded shutters for each magnetron. The grounding of the shutters (rather than floating) is inherent to the original design of the system and does not appear to affect the deposition process. as the shutters are moved away during deposition. The magnetron assemblies are situated on the bottom lid of the chamber each 90° degrees in a co-focal position with an off- axis position of 30° (Figure 2). The distance between the 50-mm diameter targets and the substrate is approximately 135 mm. When the system is in use, cleaning and bead blasting of all exposed parts of the magnetrons (i.e., parts which become coated) is performed, at least between different material systems studied, in order to avoid possible contamination. The chamber walls are water-cooled and kept near room temperature to minimize outgassing during deposition.

Figure 2 shows are schematic views of the magnetron assembly using a non-magnetic target and a magnetic target. The full descriptions of the AJA-A320 can be found in the manufacturer's manual [24]. Briefly, the magnetron AJA-A320 consists of a water-cooled copper housing containing central 10 mm diameter cylindrical NdFeB magnets (stack of three). The peripheral NdFeB magnets (stack of three) consist of a series of magnets (series of 13 magnets) positioned in a ring at 40 mm from the center of the magnetron. This configuration gives a strongly unbalanced configuration of the magnetrons which lead to high deposition rates good uniformity and dense films. The magnetic configuration can be changed to a nearly balanced magnetrons by replacing one or two magnets in the outer magnet's stacks by an iron slug. A nearly balanced configuration reduces the plasma streaming, deposition rate and film density and ideal for coatings on soft material. With a magnetic target, the center magnets are replaced by an iron slug. The close-field configuration can be changed, if necessary, by manually reversing the polarity of the permanent magnets.

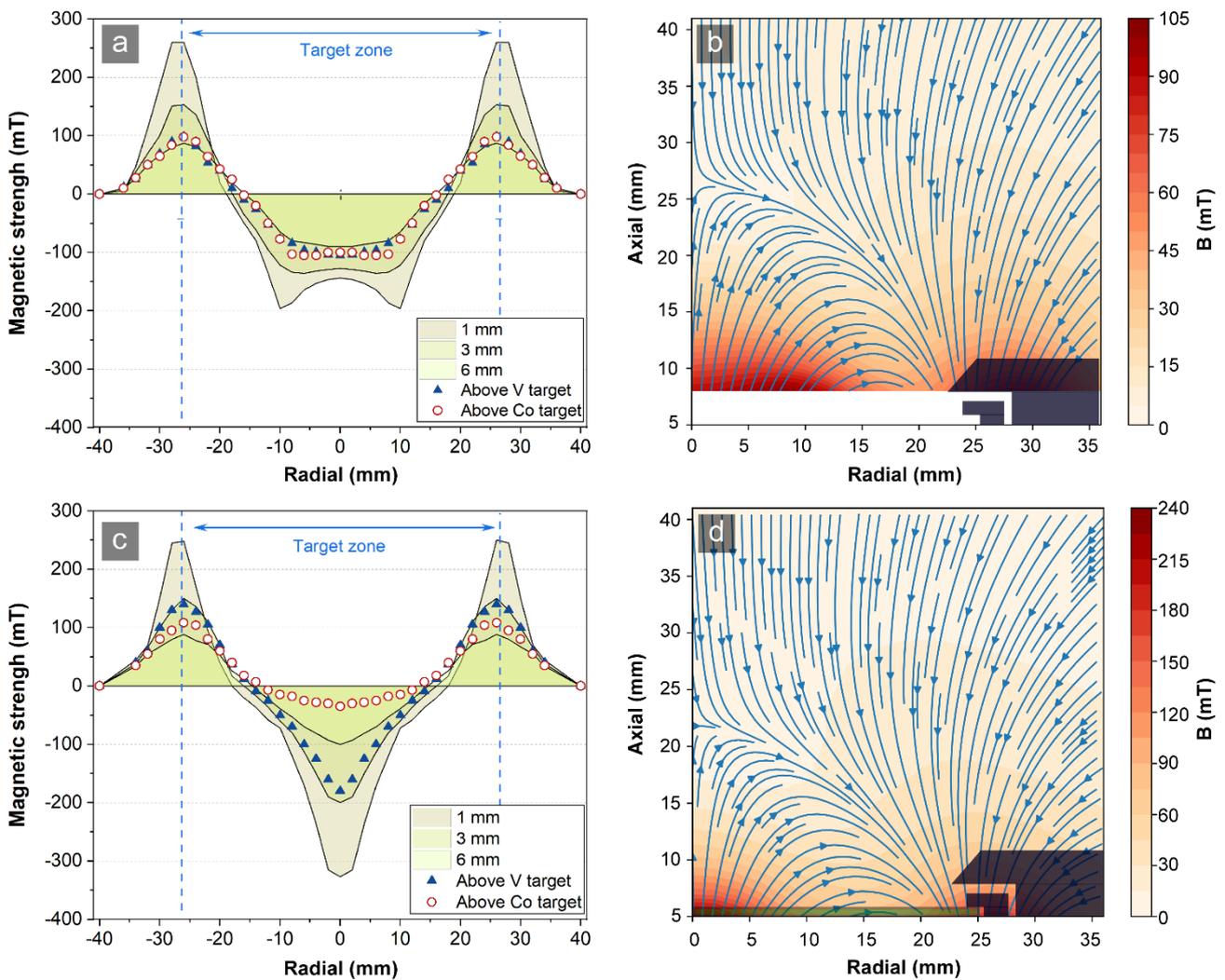

**Figure 3:** Magnetic field strength measured above the copper housing for the two configurations for: the normal component ($B_z$) of the magnetic field for a) a magnetic (i.e. Co) and c) non-magnetic target configuration, i.e. V (see figure 3). Full magnetic maps with magnetic strength and magnetic lines above the copper housing for b) a magnetic (with a cobalt target) and d) non-magnetic target configuration. Note here that the thickness of both targets is 3 mm.

Figure 3a and 3b present the magnetic field strength measured above the copper housing at different distances from the magnets in a non-magnetic and a magnetic target configuration with unbalanced magnetron. The magnetic field strength above a vanadium and a cobalt target (both 3 mm thick) are also shown. The increase of the magnetic strength is noticeable above the Co target in a magnetic configuration compared to the one in a non-magnetic configuration; the magnetic strength is increased from -30 mT to -105 mT in the center of the target. The lower values of the magnetic field strength lead to a weaker confinement of electrons near the target and consequently there may be difficulties to ignite and maintain the plasma. Note here that the maximum thickness of the target is 6 mm. The magnetic strength on top of the magnetrons was measured using a Lakeshore 420 Gaussmeter. The measured values appear smoothened, where a point maximum is expected, due to the finite size of the measurement probe area of 1x1 mm$^2$. Figure 3c and 3d shows in detail the magnetic field and lines above the copper housing of the magnetron assembly in the 2 configurations. The magnetic field of the unbalanced magnetron was measured for an individual magnetron outside the chamber and can differ when all four magnetrons are together in a close field. The magnetic strength on top of the magnetrons was measured using a Lakeshore 420 Gaussmeter attached on a xy manipulator for mapping the magnetic field strength. The unbalanced aspect of the magnetron is noticeable and can be modified by changing the strength and or the numbers of the outer ring magnets. The magnetic field strength profile gives information on the confinement of electron and ions above the target. The differences observed between the 2 configurations are important since they result in different target erosion and racetrack shape for long-time use of the target. This type of information is also important for any plasma and deposition simulation [25-27].

## 5. Power supplies

Different power sources can be connected to the magnetron assemblies. The setup is equipped with two rf power supplies (Advanced Energy RFX-600), three dc power supplies (Advanced Energy RFX-500 or RFX 1K) and three dc pulsing units (Advanced Energy Sparc-Le V). The rf units are primarily used when depositing dielectric material, especially from ceramic dielectric targets. The pulsed-dc setup (with dc power supplies connected to pulsing units) is used for depositing from metallic targets in reactive atmosphere, particularly $O_2$. The dc-units are used without pulsing for deposition of conducting materials ensuring the highest deposition rate. The rf-system uses the standard 13.56 MHz [28] sinusoidal voltage and a matching unit to minimize the reflected power. The rf-system setup results in a 50 % duty cycle and is therefore intrinsically limited in the deposition rate, compared to dc and pulsed dc.

In order to use a pulsed discharge in this system and configuration, some design considerations were needed. Most commercially available fully integrated pulsed-dc supplies are limited to high range of power (1 KW to 10 kW). Those types of high power pulsed-dc supplies are unable to deliver a well-controlled low power discharge and have a poor power resolution at low discharge power. For fundamental-research purposes using a 50-mm diameter target size, powers from 10 to 300 W (0.5 to 15 W/cm$^2$) are typically desired. To be able to deliver low pulsed powers to the magnetron, the dc power sources (Advanced Energy RFX-500) are combined to the Sparc Sparc-le V pulsing units (Advanced Energy) which then delivers a pulsed-dc discharge to the magnetron with low power and high power-resolution of 1 W from 6 W to 400 W.

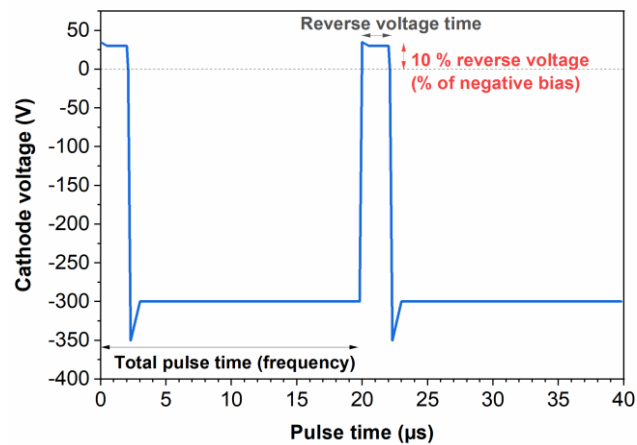

**Figure 4:** Schematic representation of a 50 kHz pulsed dc-signal with 2 µs reverse voltage time and 10 % reverse voltage magnitude from a Spar-le V unit. The overshoot becomes larger for higher pulsing frequency. For an actual waveform of this pulsing unit see reference[29].

For detailed descriptions, the reader is referred to the manual of the pulsed-dc units [24] or an article about the pulsed-dc setup with Sparc-le V [29]. In brief, the Sparc-le V units can be run in three different modes: passive, active arc and self-run mode. Passive mode (dc output) is used in target cleaning where a generated arc is slowly quenched by reverting the current upon detection of the arc. In active arc-mode, the current is more quickly diverted to stop the arc in combination with a reverted target voltage. The unit can handle up to 50 000 arcs per sec. Self-run mode works by reverting the output voltage with a set frequency between 1-100 kHz. The amplitude of the positive reverse voltage can be controlled between 10, 15 or 20% of the set negative output voltage value. The positive pulse time can be set between 1-10 µs, with fewer options the higher the set frequency is (hardware restriction). See figure 4 for description of each parameter. A crowbar delay function governs the arc quenching trial duration during which trials to clear an arc are performed through a reverse voltage before the output voltage is shorted if not successful (number of trials: 10 µs (1 try), 30 (2 tries) or 50 (3 tries) µs).

Two pulsing units, powering two magnetron assemblies, can have their pulsing frequency synchronized. In the present deposition system, when operating with two magnetrons, the bias frequency is not synchronized with the magnetrons. In summary, lower frequencies with lower positive pulse times give higher duty cycle and hence result in higher deposition rate. However, this needs to be balanced against the number of detected arcs. A higher frequency discharges clear charge built up more often, hence the likelihood for approaching arc-condition decreases. A higher frequency also results in better possibility to draw current through a dielectric material and results in higher ion current at the substrate [7] which promotes stable film growth. Therefore, often higher frequency is used for the substrate-bias pulsing than for the magnetron assembly, especially when the oxygen is fed through the wall and argon is fed at the magnetron assembly. The drawback with higher frequencies is that the output signal show larger overshoots and ripples at higher frequency [30].

## 6. Sample stage and deposition zone

The sample stage is placed at 135 mm from the targets and consists of an electrically isolated stage from the rotary shaft. Different sample holder sizes can be used up to a maximum a substrate of 50 mm (2 inches) in diameter. A Momentive Boraelectric® PBN/BN heater (R=13 $\Omega$) is situated behind the sample stage and controlled by a proportional–integral–derivative (PID: Eurotherm 3508 PID controller) controller from 100 °C to 1200 °C using a k-type thermocouple situated behind the heater. The temperature of the substrate was calibrated, using a pyrometer at the normal of the substrate (see Figure 1), using four different substrates: Ta (500 $\mu$m), Mo (500 $\mu$m), Si (500 $\mu$m) and TiN (400 nm)/c-Al$_2$O$_3$(500 $\mu$m) The substrate temperatures is comprised between 100 °C to 1100 °C for the limit range of the heater/thermocouple.

The deposition can be performed at floating potential, or with a biased or grounded substrate. The substrate bias can be applied by dc, pulsed-dc or rf voltage. An ES 0300-0.45 unit from Delta Elektronika (max: 300 V / 0.45 A) is used for the dc and pulsed dc bias voltage when combine with the Advanced Energy Sparc-Le V. An analog to digital converter on the analog readout/control connector is installed between the supply and the sample stage in order to isolate the unit properly while controlling and reading the data from that unit. This is otherwise a problem since the + and - cables needs to be flipped in order to generate a negative bias which the Elektronika unit can't handle without the extra digital converter.

A solenoid coil creating a magnetic field around the substrate can be used to increase the plasma density near substrate position. An identical model as the one presented by Engström *et al.* was adapted for this deposition chamber and placed around the sample stage [11]. The coil consists of Kapton®-insulated Cu wire (Ø=2 mm) wound ∼220 turns on a cylindrical stainless-steel frame with an inner diameter of 125 mm, resulting in a total Cu wire cross-section of ∼700 mm$^2$. A water-cooled heat exchanger was mounted in contact to the coil in its inner diameter to protect the coil from the sample stage/heater. Figure 5 shows the evolution of the magnetic field at the substrate position when a current is applied to the solenoid coil with an ES 015-10 unit from Delta Elektronika (max: 15 V / 10 A). Clear changes, observable by eye, are related to changes of ion- and electron densities directed towards the substrate in the center when a magnetic field is applied (Figure 5). The direction of the current in the coil can be altered to achieve a connected magnetic field from the sputter sources to the substrate. The strength of the magnetic field can be varied from 0 to 12 mT (120 Gauss) while the lateral intensity of the plasma is reduced for the higher magnetic strength as observable already for a plasma running with a coil at 7 A for example (figure 5).

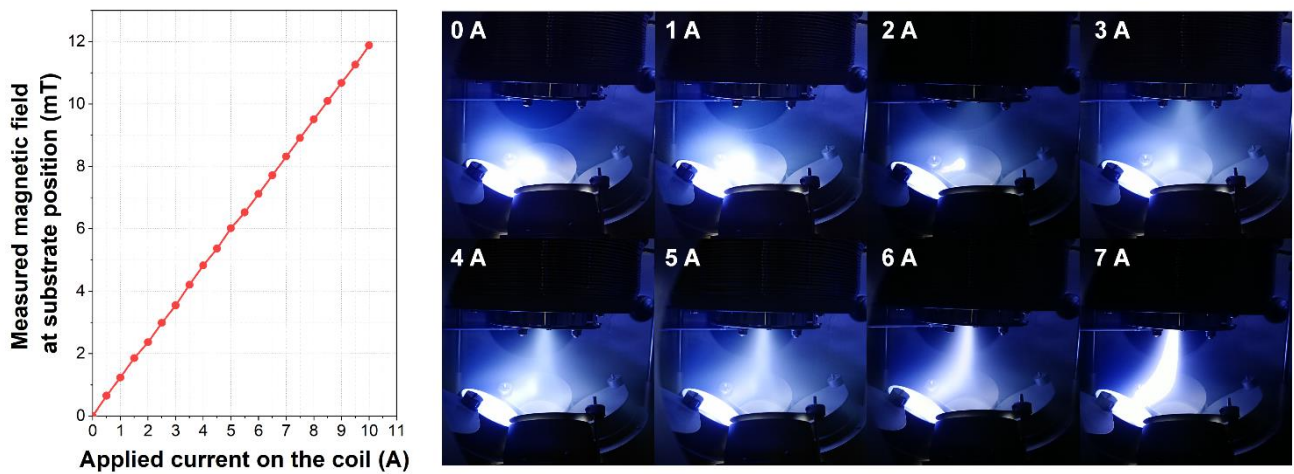

**Figure 5:** Evolution of the magnetic field at the substrate position versus the applied current to the solenoid coil. Visual evolution of the plasma for different applied current (Target: cobalt / power: 100 W (dc) / pressure: 0.4 Pa (3 mtorr) Ar).

SIMTRA (version 2.2.0) software [31] was used to simulate and estimate the characteristics of the particles reaching the deposition zone (substrate). Figure 6 shows an example of the density of particles reaching the deposition zone and their average energies with and without rotation of the substrate (no substrate bias and no coil). The simulation was performed using aluminum as material target, a Thompson energy distribution with a maximum energy of 380 eV, a surface energy of 3.36 eV, an axial-symmetric racetrack on the target provided in SIMTRA software package, the screened Coulomb potential described with a Molière screening function [32], a total number of particles sputtered of $3.6\times10^7$ at pressure of Ar of 0.4 Pa. The SIMTRA software does not provide the option of rotation of the circular dummy object (substrate) during simulation. In order to visualize the case with rotation of the substrate, several individual simulations of $1\times10^6$ sputtered particles were performed each 10 degree of rotation of the dummy object and later computed to one single picture with $3.6\times10^7$ sputtered particles in total.

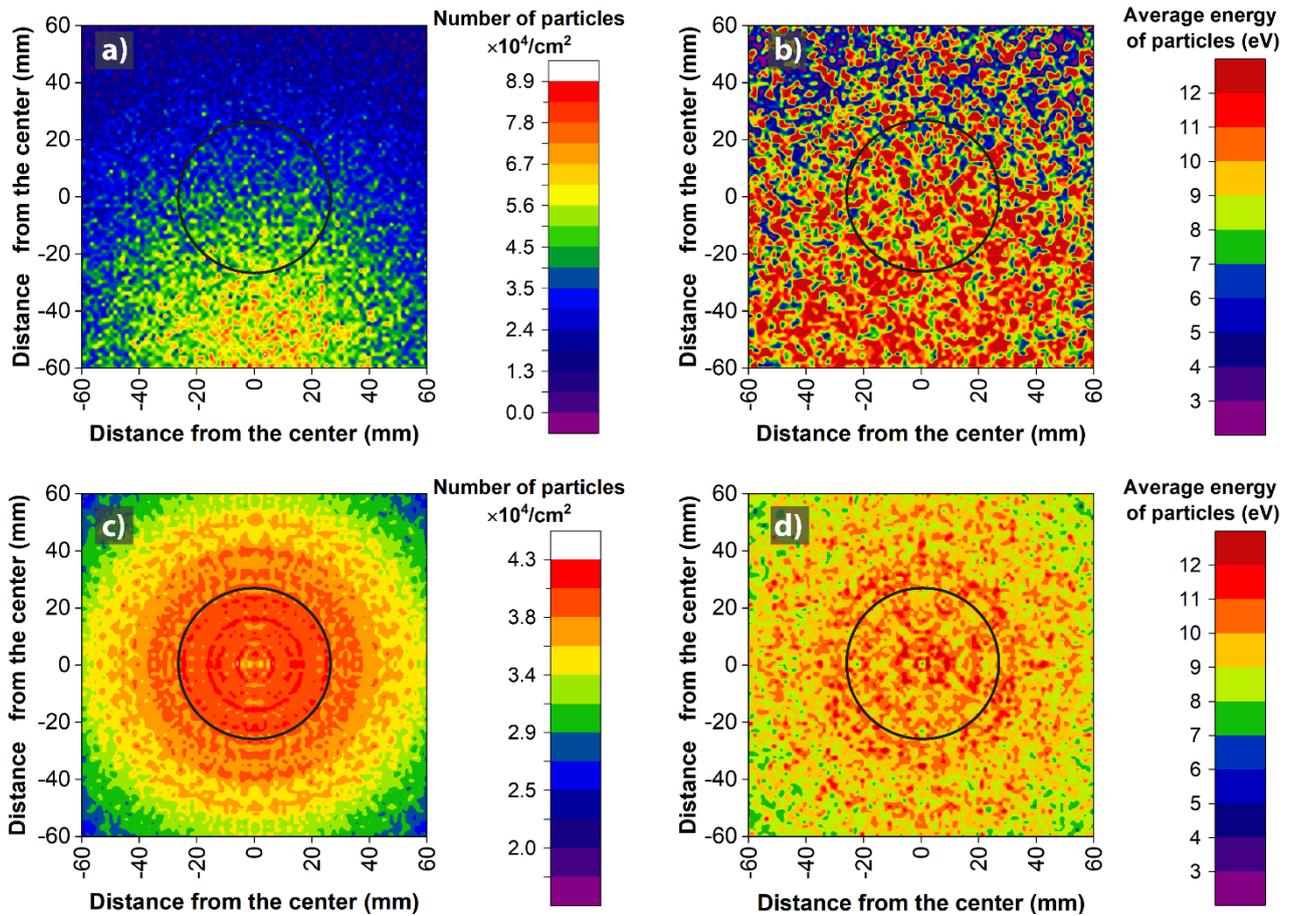

**Figure 6**: SIMTRA simulation: (a and c) number of aluminum atoms deposited and (b and d) their energy at 0.4 Pa Ar pressure on a dummy object (substrate) placed at the deposition zone. Simulation performed (a and b) at fixe position (no rotation) with the magnetron directed towards the minus y-scale and (c and d) with computed rotational substrate. The black circle represents the maximum substrate size (50 mm in diameter, 2 inch in diameter).

The configuration of the deposition chamber and the tilting of the magnetron assembly do create inhomogeneity of the deposition at the substrate zone. Without rotation of the substrate, the number of particles per centimeter square and their average energies are greater towards the direction of the magnetron (figure 6a) and 6b). The number of particles seems to be very dependent of the position of the magnetron with a variation of 50 % from $6.7 \times 10^4$ (-30 mm) to $3.0 \times 10^4$ (+30 mm) particles/cm$^2$. The average energy of the particles is less affected with a variation from 9-11 eV (-30 mm) to 7-9 eV (+30 mm). The rotation of the substrate averages the number of particles and gives a homogenized zone between -30 to +30 mm. In the deposition zone (-25 to +25), the number of particles is estimated to vary between $4 \times 10^4$ and $4.4 \times 10^4$ particles/cm$^2$ with an energy between 7-12 eV.

# 7. LabVIEW software control

Each unit is connected to a National Instruments™ DAQ card directly connected to computer running a custom-made LabVIEW software which enables the control over all the different commands and process parameters. Three different cards from National instruments are mounted in a CompactDAQ chassis: 1) one NI6264 with 16 channels analog outputs -10 to + 10V (to control the instruments where the output is controlled by analog signals); One NI9403 with 32 channels digital/inputs outputs (to control the digital input/outputs of equipment); 3) One NI9205 with 32 analog inputs (to measure analog data). Other types of connection are used such as USB and Rs232 for controlling the Eurotherm PID controller or the Pulsed DC units from Advanced Energy. The LabVIEW code is provided as Supplementary Information and may be freely used and modified to fit any other deposition system, on condition that the present article is credited as source and modifications are specified.

Figure 7 shows the different user windows of the software. The first window consists of the control modules where each controllable parameter can be changed, programed, and saved for reproducibility. The second window consists of a visualization window for direct observation of variables during deposition. The different parameters measured before, during and after deposition can be directly observed in a real-time plot with up to 6 plots being active simultaneously. The program gives the possibility to record and save all the parameters (pressures, gas-flows, powers, currents, voltages, temperatures, arc counts, time) of a deposition. The software can automatically control most aspects of the system, for example, different options are available such as a target sputter cleaning procedure and automated time control of the shutters. The automated control of the shutters allows the deposition of more complex structures, such as multilayers or superlattices, where deposition time, bias, and coil characteristics can be changed individually. To assure a good synchronization response between the different shutters, the substrate bias and coil power supplies, a systematic "waiting time" command of 1 second is added to the LabVIEW code between the command to close (or open) the shutter(s) and the command to open (or to close) the other shutter(s). The implementation of the superlattice is done using a comma-separated values file (csv file). where each row corresponds to the layer of the superlattice and each column corresponds to the different parameters: the deposition time; the position of the shutter for each magnetron assembly (open (1)/closed (0)); the substrate bias (on (1) / off (0)) and its values; and finally, the coil (on (1) / off (0)) and its parameter values. The use of a csv file gives an unlimited possibility for governing the deposition with multiple combinations of parameter and infinite numbers of layers. More information about the structure of the software and a brief description on how to implement and modify it can be found in the Supplementary Information.

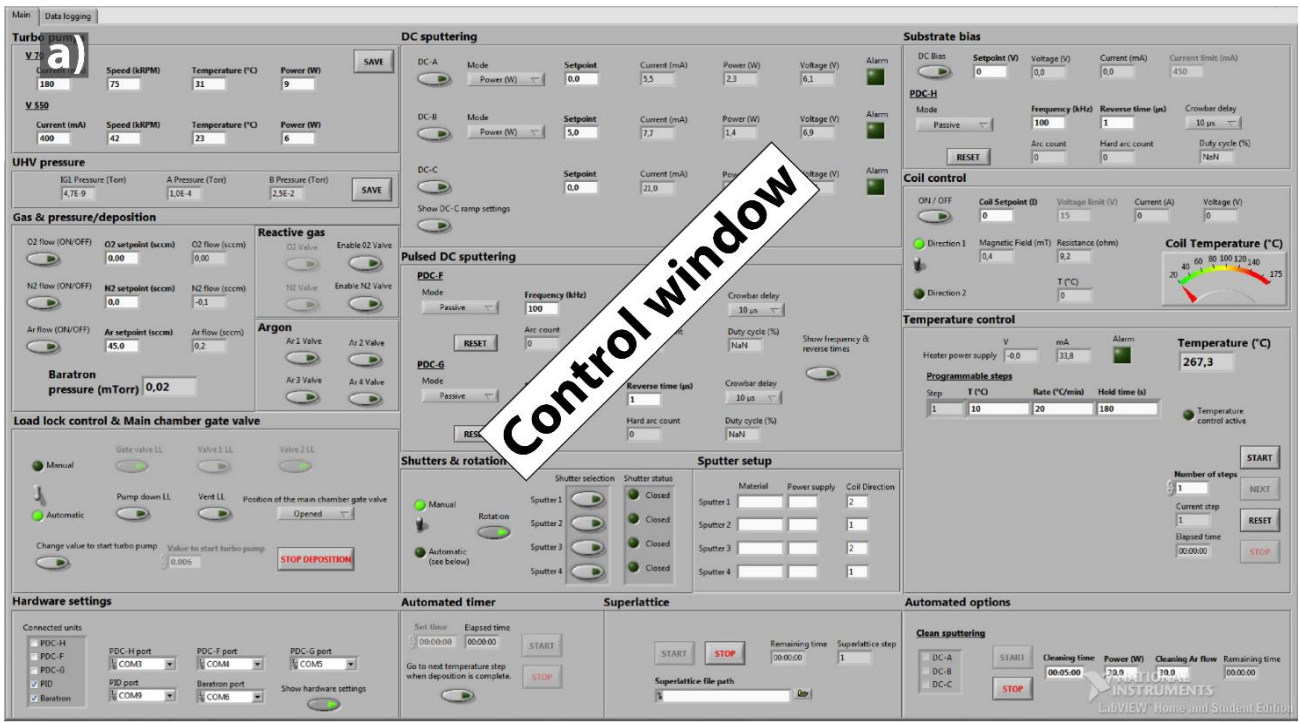

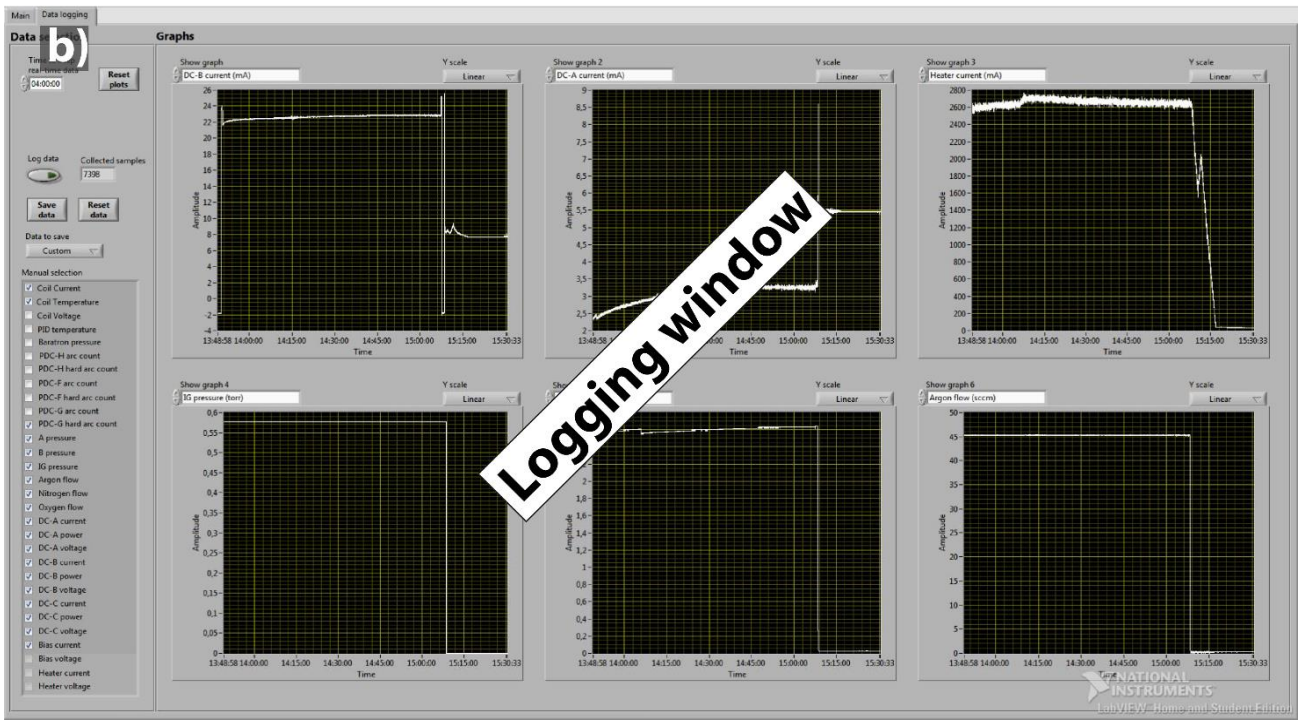

**Figure 7:** Screen windows of the LabVIEW program for control for the deposition chamber. One windows for controlling the different parameters (a) and one for logging the different variables during deposition which can be saves into a data log file (b).

## 8. Concluding remarks

The present article has provided a detailed description of the design of a lab-scale deposition chamber for magnetron sputtering adapted for the versatility required for the use for the deposition of metallic, oxide, nitride and oxynitride films combined with largely computerized control and logging. Custom-made LabVIEW software enables the control over all the different commands and process parameters. The LabVIEW code is provided as Supplementary Information and may be freely used and modified to fit any other deposition system, on condition that the present article is credited as source and modifications are specified.

## Acknowledgements

The authors would like to acknowledge Dr. Michal Zanaska from the Department of Physics, Chemistry and Biology (IFM) at Linköping University for his help for the measurement of the magnetic field above the magnetrons. The authors acknowledge the Swedish Research Council (VR) under Project No. 2016-03365 and 621-212-4368, the Swedish Energy Agency under Project 46519-1, the Knut and Alice Wallenberg Foundation through the Wallenberg Academy Fellows program (KAW-2020.0196), and the Swedish Government Strategic Research Area in Materials Science on Functional Materials at Linköping University (Faculty Grant SFO-Mat-LiU No. 2009 00971).